\documentclass[useAMS,usenatbib, usegraphicx]{mn2e}
\usepackage{bethmacros}
\usepackage{amssymb}
\usepackage[T1]{fontenc}
\usepackage{aecompl}

\def\spose#1{\hbox to 0pt{#1\hss}}
\def\lta{\mathrel{\spose{\lower 3pt\hbox{$\mathchar"218$}}
\raise 2.0pt\hbox{$\mathchar"13C$}}}
\def\gta{\mathrel{\spose{\lower 3pt\hbox{$\mathchar"218$}}
\raise 2.0pt\hbox{$\mathchar"13E$}}}

\title[Superbubble Feedback Model]{A Superbubble Feedback Model for Galaxy Simulations}

\author[Keller et al.]{B.\,W. Keller$^{1}$\thanks{Email: kellerbw `at'
    mcmaster.ca},  J. Wadsley$^{1}$, S.M. Benincasa$^{1}$,  H. M. P. Couchman$^1$
    \vspace*{6pt}\\
    $^1$Department of Physics and Astronomy, McMaster University, Hamilton, Ontario, L8S 4M1, Canada
}

\begin{document}
\maketitle
\label{firstpage}

\begin{abstract}
We present a new stellar feedback model that reproduces superbubbles.
Superbubbles from clustered young stars evolve quite differently to individual
supernovae and are substantially more efficient at generating gas motions.  The
essential new components of the model are thermal conduction, sub-grid
evaporation and a sub-grid multi-phase treatment for cases where the simulation
mass resolution is insufficient to model the early stages of the superbubble.
The multi-phase stage is short compared to superbubble lifetimes.  Thermal
conduction physically regulates the hot gas mass without requiring a free
parameter.  Accurately following the hot component naturally avoids overcooling.
Prior approaches tend to heat too much mass, leaving the hot ISM below $10^6$ K
and susceptible to rapid cooling unless ad-hoc fixes were used.  The hot
phase also allows feedback energy to correctly accumulate from multiple,
clustered sources, including stellar winds and supernovae.

We employ high-resolution simulations of a single star cluster to show the model
is insensitive to numerical resolution, unresolved ISM structure and suppression
of conduction by magnetic fields.  We also simulate a Milky Way analog and a
dwarf galaxy.  Both galaxies show regulated star formation and produce 
strong outflows.
\end{abstract}


\section{Introduction}
Galaxies are star factories:  with their large potential wells, they accrete gas
and convert that gas into stars.  The throttle for this process
is the energy released from these stars through winds and supernovae: stellar
feedback.  Without this large source of energy ($\sim 3\times 10^{38}$ erg
s$^{-1}$ per solar mass of stars) to stir and heat the interstellar medium, star
formation would consume all available gas for every galaxy in less than a Hubble
time.  Not only does stellar feedback allow star formation to self-regulate, it
is one of the most important processes in producing a multiphase ISM
\citep{McKee1977}.  A third way feedback shapes the history and structure of gas
in a galaxy is by cycling (and even ejecting) gas through outflows.  Galactic
winds can remove potential star-forming gas from a galaxy by propelling it out
of a galaxy faster than the escape velocity.  Gas `fountains' can reduce the
cold gas available in a galactic disc by cycling it between the disc and high in
the galactic halo. This acts to temporarily store gas in a reservoir above the
galactic plane, where it is too hot and diffuse to form stars.  These outflows
are likely an important component in determining the ultimate fate of a galaxy,
and are the most plausible mechanism for metal enrichment observed in the
circumgalactic medium \citep{Songaila1996,Dave1998}.

Much work has been done to build feedback models based on the evolution of
individual supernova blastwaves \citep[e.g.][]{Stinson2006}, but these efforts
have overlooked two key factors.  First, star formation is clustered; new stars
are spatially and temporally correlated, and feedback from their individual
winds and supernovae merge, thermalize and grow as a {\it superbubble} rather
than a series of isolated supernovae.  Second, because superbubbles have both
hot gas $>10^6\;\mathrm{K}$ and sharp temperature gradients, thermal conduction
is significant \citep{Weaver1977}.
Omitting this process can cause one to
incorrectly estimate the interior density (and thus the amount of hot gas) of
superbubbles by orders of magnitude, regardless of whether or not one can
resolve the superbubble.  In simulations of galaxy evolution, the temporal
resolution required to resolve the pre-thermalization Sedov phase is out of
reach (on the order of $100\;\mathrm{yr}$), and even the post-thermalization
early superbubble can require shorter timesteps than are practical. Worse still,
during this period the amount of mass contained within the hot, rarefied
interior of a superbubble is less than the mass of the progenitor star cluster.
This can make it impossible to spatially resolve this stage in simulations where
resolution elements are comparable in mass to star particles.  This leads to
denser, cooler feedback regions.  These overcool and lead to ineffective
feedback overall \citep{Katz1992}.

A number of approaches exist to attack the problem of overcooling.  The earliest
techniques were to simply deposit a fraction of the energy released in feedback
events as kinetic energy (\citet{Navarro1993}, \citet{Mihos1994},
\citet{Dubois2008}, etc.).  \citet{Gerritsen1997} detailed a second approach; by
introducing cooling shutoff, where feedback-heated gas is explicitly prevented
from cooling radiatively, and \citet{Thacker2000} explored a range of different
times for this shutoff period.  \citet{Stinson2006} proposed using the time
required to resolve a Sedov-Taylor blastwave, and showed that this can be an
effective way of modelling feedback in cosmological simulations of galaxy
evolution.  \citet{Agertz2013} used a decaying non-cooling energy, where energy
in a non-cooling state decayed back to the `normal' cooling form. Another
technique is to manually decouple density estimates and hydrodynamic
interactions between feedback-heated gas and the cold ISM (\citet{Marri2003},
\citet{Scannapieco2006}, etc.).  
With extremely high resolution, it
is possible to generate rarefied hot gas from feedback directly without 
a subgrid treatment \citep[e.g.][]{Hopkins2012}.  
However, as we argue here, mass transfer between hot and cold gas
depends on the physics of conduction which relies on sub-parsec
gradients.  These are beyond the reach of even the highest resolution 
galaxy scale simulations so some subgrid modeling may be unavoidable.

Another approach has been to explicitly model the multiphase ISM
below the resolution limit.  \citet{Springel2003} described a multiphase
model based on the theoretical framework of \citet{McKee1977}.  Each
particle was composed of an isobaric mix of cold clouds and ambient warm
to hot gas.  Radiative cooling coverts warmer gas into cold.  The cold
phase forms stars on a characteristic timescale chosen to yield a
Schmidt-type star formation law.  An empirical model of stellar feedback
evaporates the cold phase.  \citet{Springel2003} recognized that while
this model works well for simulating star formation and feedback in
quiescent galaxies, the coupling of hot and cold mass can prevent hot
gas from leaving the disc as winds or outflows.  Their solution was to
convert a fraction of the feedback energy in a kinetic kick on selected
particles, in the same vein as \citet{Mihos1994}.  

Both \citet{DallaVecchia2012} and \cite{Hopkins2012} have shown
that it is possible to get consistent wind results for a given energy input model.   
\citet{DallaVecchia2012} demonstrated that simply depositing 
 energy stochastically to ensure a constant temperature increase for 
feedback-heated gas can directly generate winds.    
They found that for the same feedback energy,
changing the temperature of feedback-heated gas results in significant
differences in both star formation regulation and galactic outflows.  Higher
feedback temperatures, $\Delta T > 10^7\;\mathrm{K}$, avoid overcooling and
allow for more efficient regulation and higher velocity galactic winds.  
This still leaves open the question of what sets this temperature?  
This question is equivalent to asking: {\it what sets the mass-loading in stellar
feedback?}   Previous feedback models have not explored key
physical mechanisms, such as conduction, that affect mass-loading.

The above sub-grid models are reasonably successful at preventing
overcooling.  However, many of them have limitations which are
increasingly severe with improving resolution.  For example, stellar feedback
in the form of kinetic energy is rapidly converted into thermal energy as it
encounters the ISM and shocks \citep{Durier2012}.  In nature,
the gas heated by feedback doesn't completely stop cooling radiatively, it
merely cools inefficiently.  Applying a cooling shutoff is unlikely to give the
correct behaviour in different star forming environments and is also dependent
on the integrated energy injection from all nearby stars.  Finally, because
star formation is clustered, feedback is localized within starforming regions. 

Recent studies such as \citet{Nath2013} and \citet{Sharma2014} have emphasized
that feedback from clustered stars forms superbubbles, which behave quite
differently from isolated supernovae.   A key outcome is that superbubbles are
intrinsically more efficient than individual supernovae at converting feedback
energy into gas motions, particularly at late times and over larger scales.  The
reason for this difference is that gas heated by feedback remains distinct from
the cooler surrounding material.  Most current models smear together the hot
bubble with the cold shell surrounding it.  This results in an intermediate
effective temperature that is prone to overcooling.   Separating the hot and
cold phases automatically avoids overcooling.  \citet{DallaVecchia2012} achieved
this with a stochastic feedback model.  An alternative approach is to add an
explicit hot reservoir to accumulate feedback mass and energy.  This avoids
overcooling without artificially turning off cooling and correctly handles
feedback from multiple sources over time without resorting to a stochastic
approach.  Such a model still leaves the bubble mass as a free parameter.
\citet{MacLow1988} showed that thermal conduction controls the mass flow into
the hot bubble from the cold shell.  This evaporation process regulates the
temperature of the hot bubble and determines how much mass is heated by
feedback.   Adding a sub-grid model for evaporation allows the physics of
thermal conduction to set bubble temperatures and masses.

Drawing on these facts, we can construct a {\it superbubble}-based feedback
model.  As outlined in \citet{MacLow1988}, superbubbles efficiently convert
feedback energy into thermal energy in a hot phase and kinetic energy in an
expanding cold shell.  The rarefied hot phase cools inefficiently, avoiding
overcooling.  Thus a correct model requires following distinct hot and cold
phases even when they may be difficult to resolve directly.  A new requirement
with respect to prior feedback models is the inclusion of thermal conduction.
Conduction both smooths the temperature distribution in the hot gas and mediates
mass flows where hot gas meets a cold phase.  Thus a second feature of such a
model is that an explicit physical process sets the amount of hot gas in the ISM
and in outflows.

In this paper, we begin by explaining the theoretical underpinning and
numerical implementation of the superbubble-based feedback method in section~\ref{method}. 
In section~\ref{cluster} we use a single star cluster to illustrate the
effectiveness of our model at capturing the basic behaviour of superbubbles at
high and low resolution.  In section~\ref{galaxy} we apply the model to
simulations of isolated galaxies to explore the impact on the galaxy
scale ISM and the production of outflows.  

\begin{figure*}
    \includegraphics[natwidth=1833, natheight=331, width=\textwidth]{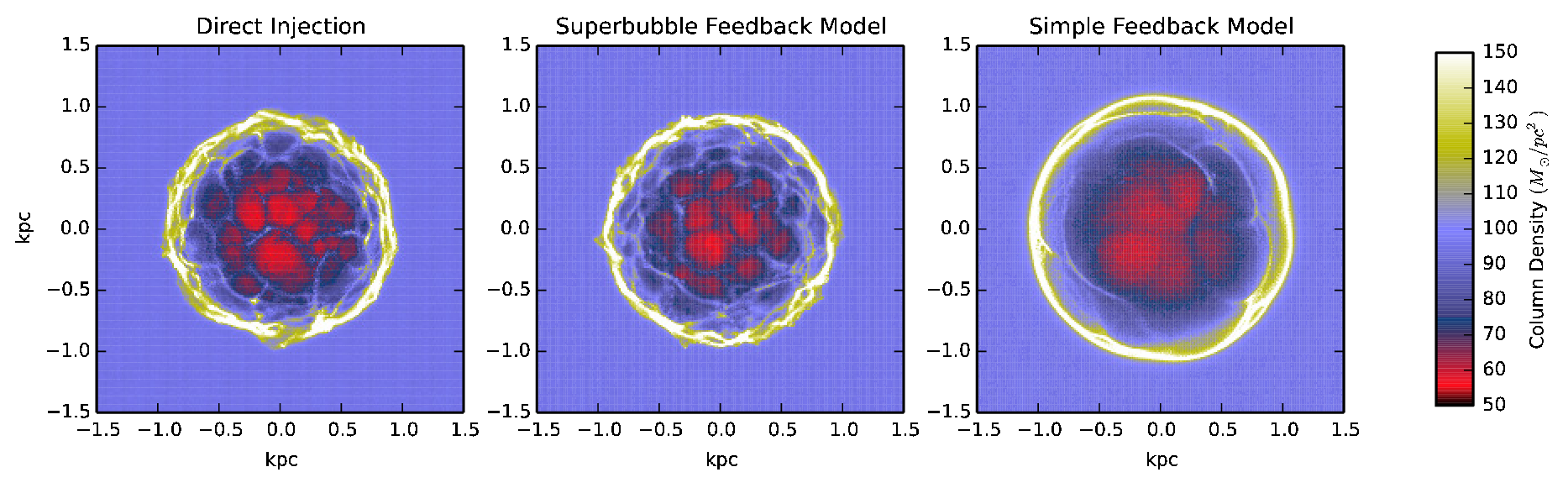}
    \caption{Column density projections from the simulations of a single star
        cluster with mass $3\times10^4\;\mathrm{M_\odot}$, $50\;\mathrm{Myr}$
        after the cluster has formed.  The superbubble feedback model (center
        column) produces bubbles with radius and enclosed mass that match well
        to the direct injection simulations (left column).  This is shown
        quantitatively in figure~\ref{onestar_homogeneous} and
        figure~\ref{onestar_clumpy}.  The simple model (right column), despite
        injecting the same amount of energy, fails to generate enough hot mass
        inside the bubble, and subsequently suppresses the growth of Vishniac
        instabilities along the bubble edge because of the poorly resolved hot
        interior.}
    \label{onestar_column}
\end{figure*}

\section{Thermal Conduction and Feedback Method}\label{method}
Our new treatment of feedback has three components.  The first is the addition
of thermal conduction.   Inside resolved hot bubbles, thermal conduction
maintains uniform temperatures.   In the presence of strong gradients, thermal
conduction can lead to evaporative mass flows from cold to hot gas.  The second
component is a stochastic model of evaporation to allow resolved hot gas to
continue to gain mass from nearby cold gas.  Thus, the amount of cold gas heated
by feedback is not a free parameter, but is set by thermal conduction.  
Without a mechanism like thermal conduction, this hot gas mass is set by
how many fluid elements have feedback energy deposited into them.   It is
important to note that in our model these processes operate everywhere
temperatures are above $10^5K$. 

Finally, in the first few $\mathrm{Myr}$ of feedback heating, the mass contained
within a hot bubble can be smaller than the simulation gas mass resolution.  To
prevent overcooling, we allow resolution elements to become briefly two-phase; a
hot interior (bubble) in contact with a cold shell.  Evaporation of the cold
shell moderates this two-phase period, rapidly returning particles to single
phase once their cold phase has been fully evaporated.  Our model does not
assume all fluid elements are multiphase, but only those in which a partial
feedback region exists.  This allows the model to follow winds and outflows
without continuing to rely on sub-grid machinery.  

Young stellar population steadily release energy in the form of winds and SNe at
a rate of $3\times 10^{38}\;\mathrm{erg\; s^{-1} M_\odot^{-1}}$ for around 40
Myr~\citep{Leitherer1999}. We deposit this energy into the gas particle nearest
to the star particle.  In following with past work, we use the feedback rates
and times for supernovae, but the method is general enough to handle heating
from stellar winds, ionization, etc.  Heating takes effect
$4\;\mathrm{Myr}$ after a star forms, and continues until $30\;\mathrm{Myr}$
after the star particle forms (the time associated with SNII from OB stars).  Thus,
each supernova releases $10^{51}\;\mathrm{erg}$, and each star particle will
release $10^{49}\;\mathrm{erg M^{-1}_\odot}$ using the \citet{Chabrier2003} IMF.

\subsection{Thermal Conduction}
In an ionized gas, thermal conduction, mediated by electrons,
transports heat down temperature gradients.  This flux, $\mathbf{Q} =
-\kappa\nabla T$, depends on the temperature gradient and the conduction
coefficient, as derived by \citet{Cowie1977},
\begin{equation} \label{coefficient}
    \kappa(T) = 1.8\left(\frac{2}{\upi}\right)^{3/2}\frac{T^{5/2}k_B^{7/2}}{m_e\,e^4\ln\Lambda}.
\end{equation}
This coefficient depends only on the Coulomb logarithm $\ln\Lambda$ (which has
an extremely weak dependence on density), and is well approximated by $\kappa(T)
= \kappa_0\,T^{5/2}$, where $\kappa_0$ is $6.1\times10^{-7}\;\mathrm{erg\;s^{-1}
K^{-7/2}cm^{-1}}$ in the absence of magnetic fields.  In order for this
situation to not cause spontaneous currents, a corresponding mass flux,
{\it in the opposite direction}, must also occur \citep{Cowie1977}.  With
spherical symmetry, this implies a mass flow rate that depends on the sound
speed in the hot gas, $c_{\rm s}$, as follows,
\begin{equation}\label{conservation}
    \frac{5}{2}\dot M c_{\rm s}^2 = 4\,\upi^2r^2\kappa(T)\frac{{\rm d}T}{{\rm d}r}.
\end{equation}

These rates hold only when the mean free path of electrons in the medium is
smaller than the scale length of the temperature gradient.  If the gradient
becomes steep enough, the heat flux (and corresponding mass flux) saturates at a
value that depends only on the density, temperature, and thermal velocity of the
electrons in the medium \citep{Cowie1977},
\begin{equation}\label{heatflux}
    \mathbf{Q} = \nabla\left(\frac{3}{2}n_ek_BT_ev_e\right).
\end{equation}
This saturation has the convenient numerical side effect of setting the smallest
timestep required to resolve this mass flux to $\sim 1/17$ the standard Courant
time.

In situations where the temperature gradient is embedded in a strong magnetic
field, the value of $\kappa_0$  can be reduced by factors approaching an order
of magnitude depending on the strength and configuration of the magnetic fields
\citep{Cowie1977}.

The edge of a feedback-driven superbubble presents a strong discontinuity in
both temperature and density.  Interior to the bubble, gas has temperatures of
$\sim 10^6\;\mathrm{K}$ and densities of $\sim10^{-3}\;\mathrm{cm^{-3}}$, while
the shell can have temperatures below $100\;\mathrm{K}$ and densities of $\sim
10\;\mathrm{cm^{-3}}$ \citep{Chevalier1974}.  This generates significant mass
and energy flows due to thermal conduction.

\subsection{Evaporation}
The dominant physical process governing mass flux between the hot and cold
phases of a feedback bubble is thermal conduction between the dense shell and
the hot interior.   As this process takes place on length scales far below the
resolution limit, we must capture its effects in a subgrid model.  For the thin
shell surrounding a feedback bubble, thermal conduction causes an evaporative
mass flux from the cold shell into the hot bubble.  Following
\citet{MacLow1988}, the mass flux into a bubble with interior temperature $T$
is,
\begin{equation}\label{massflux}
    \frac{{\rm d }M_b}{{\rm d}t} = \frac{4\,\upi\mu}{25k_B}\kappa_0\,\frac{\Delta T^{5/2}}{\Delta x}\ A,
\end{equation}
where $ A$ is the bubble surface area and $\Delta x$ is the thickness of
the hot layer.  For the Smoothed Particle Hydrodynamics (SPH) method used for
our tests, we calculate evaporation using the outer layer of hot particles
bordering the cold gas.  The members of this layer are those with no other hot
particles within $45^o$ of the vector to the centre of mass of their cold
neighbours.  As we cannot directly determine the radius of a well-resolved
superbubble without expensive non-local calculations, we need a way to allow
each particle to estimate the radius and thus its fractional contribution of the
shell surface area.  The total area estimated by all hot particles in a bubble
should approach $\sim 4\upi R^2$ where $R$ is the bubble radius.  For a poorly
resolved bubble $R \sim 1-2\,h$, where $h$ is a hot particle's SPH smoothing
length.  For larger bubbles each hot particle contributes an area of $\sim h^2$
and each particle sees a nearly plane-parallel section of the cold shell.  We
examined a number of bubbles at different stages of growth, as well as a
plane-parallel slab, and empirically found that a good per-particle area
estimate was $A = \frac{6\,\upi\,h^2}{N_{\rm hot}}$, where $N_{\rm
hot}$ is the number of hot neighbours for that particle and $\Delta x = h$.  We
stress that this is a fit for our specific SPH neighbour approach (described
below) that should be recalibrated for other codes.

The mass evaporation rate can be converted into a probability that a resolution
element with cold mass $m$ converts into a hot one over a time
period $\Delta t$ as follows, 
\begin{equation}\label{evaporation_probability}
    P_{evap} = \frac{{\rm d}M_b}{{\rm d}t} \frac{\Delta t}{m}.
\end{equation}
This allows us to stochastically choose full particles to evaporate, and prevent
overcooling due to fractional particle evaporation (exactly the same overcooling
problem seen in feedback heating).  Each hot particle determines how many
cold-shell particles $N_{evap}$ will evaporate each timestep, and then chooses
the $N_{evap}$ nearest cold particles, and averages the thermal energy of itself
and those particles.  These particles spontaneously join the hot bubble
(demonstrated in the figures in section~\ref{clumpy}).  The thermal conduction
rates calculated are capped using the saturation values derived in
\citet{Cowie1977}.  Within an SPH framework, this is well approximated using,
\begin{equation}\label{saturation_rate}
    \frac{{\rm d}M_{sat}}{{\rm d}t} = 17\rho\,c_{\rm s} h^2.
\end{equation}

\begin{figure*}
    \includegraphics[natwidth=1833, natheight=631, width=\textwidth]{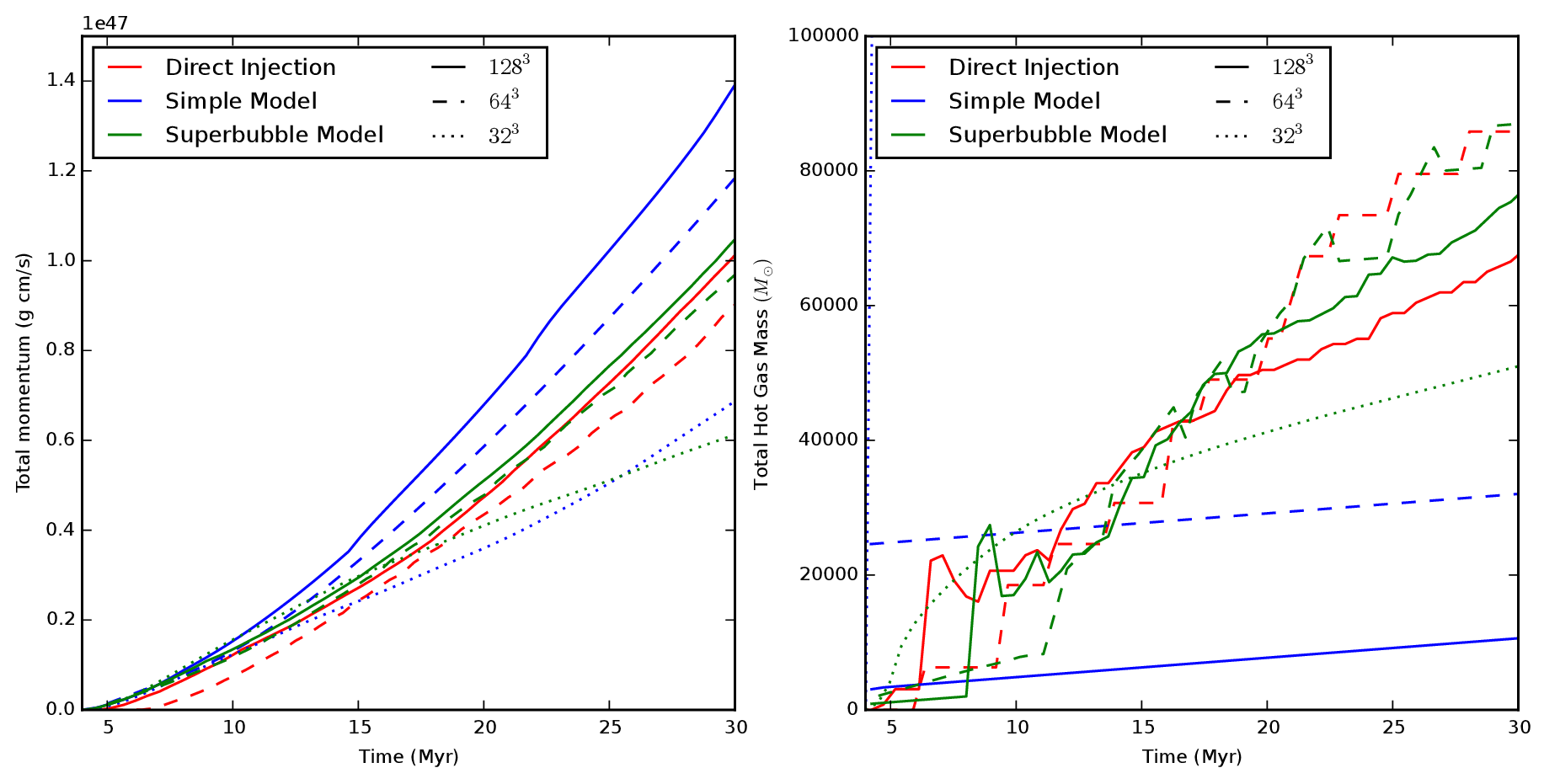}
    \caption{Feedback effects as a function of resolution.  The above figure
        shows on the left the total radial momentum imparted to the medium, and
        on the right the total amount of mass heated to above
        $10^5\;\mathrm{K}$.  The red curves are the direct injection results, the
        blue corresponds to the simple feedback, while the green shows the
        results of the superbubble model.  Solid lines show the values from
        simulations run with resolutions of $128^3$ particles, dashed for
    $64^3$, and dotted for $32^3$.
Note that for the $32^3$ run, the simple model produces $\sim 2\times
10^5\;M_\odot$ of hot gas.}
    \label{onestar_homogeneous}
\end{figure*}

\subsection{Multiphase Fluid Elements}
When feedback energy is deposited in a low-resolution simulation, or is
deposited as a luminosity, the temperature and density are certain to be under-
and over-estimated respectively.  This problem leads to the well-known
overcooling problem that has typically been addressed by disabling
cooling for some amount of time for feedback-heated gas (e.g.
\citet{Stinson2006}, \citet{Springel2003}, or by stochastic feedback heating,
where the temperature change of a fluid element heated by feedback is fixed to a
constant value (e.g.  \citet{DallaVecchia2012}.  We employ a third option:
storing feedback energy in a second phase that is in pressure equilibrium with
the rest of the fluid inside an element.

Fluid elements (gas cells or particles) enter the multi-phase state if they are
given energy from feedback, and if their temperature is below $10^5\;\mathrm{K}$
(the `hot' threshold).  Multiphase elements have two values for their mass and
energy, related to their total mass $m$ and energy $E$ by:

\begin{equation}\label{multiphase_fractions}
    m = m_{\rm hot}+m_{\rm cold}\\
    E = u_{\rm hot}m_{\rm hot}+u_{\rm cold}m_{\rm cold}
\end{equation}

Assuming pressure equilibrium, both phases will have the same pressure $P$, and
their densities are found using this and the total density $\rho$:
\begin{equation}\label{multiphase_pressurebalance}
    \frac{P}{\rho} = \frac{(\gamma-1)E}{m}
\end{equation}
\begin{equation}\label{multiphase_densities}
    \rho_{\rm cold} = \frac{P}{(\gamma-1)u_{\rm cold}}\\
    \rho_{\rm hot} = \frac{P}{(\gamma-1)u_{\rm hot}}
\end{equation}

Both the cold and hot phases are allowed to radiatively cool using their
separate temperatures and densities.   When $PdV$ work is done to a multiphase
particle, it is shared between the two phases weighted by their respective
fraction of the total energy E,
\begin{equation}\label{multiphase_PdV}
    \dot u_{\rm PdV,cold} = m\ \dot u_{\rm PdV}\frac{u_{\rm cold}}{E}\\
\end{equation}
\begin{equation}
    \dot u_{\rm PdV,hot} = m\ \dot u_{\rm PdV}\frac{u_{\rm hot}}{E}.
\end{equation}
In the absence of heating and cooling, this allows each phase to correctly
maintain constant entropy as the densities change.

Mass flux between the hot and cold phase is calculated in a continuous manner
consistent with the prior evaporation scheme.  Each timestep, a multi-phase
element evaporates a fraction of its cold phase into the hot phase,

\begin{equation}\label{multiphase_evaporation}
    \frac{{\rm d}M_b}{{\rm d}t} = \frac{16\upi\mu}{25k_B} \kappa_0 (T_{\rm hot}^{5/2})\ h.
\end{equation}

Once an element has evaporated all of its mass into the hot phase, or if the hot
phase cools below $10^5\;\mathrm{K}$, it is returned to the single-phase state.

\subsection{SPH Implementation}

We implemented this method in the SPH code {\sc GASOLINE}
\citep{Wadsley2004} with updates described in \cite{Shen2010}.  These
include a sub-grid model for turbulent mixing of metals and energy.  The
heating and cooling include photoelectric heating of dust grains, UV heating
and ionization and cooling due to hydrogen, helium and metals.

The SPH hydrodynamic treatment has had some further, key updates.  We currently
use a standard SPH density estimator but a  geometric density average in the SPH
force expression: $(P_i+P_j)/(\rho_i\,\rho_j)$ in place of
$P_i/\rho_i^2+P_j/\rho_j^2$ where $P_i$ and $\rho_i$ are particle pressures and
densities respectively. This force expression alleviates numerical surface
tension associated with density discontinuities, which is important for correct
Kelvin-Hemholtz instabilities and ablation of cold blobs (as in the Agertz et al
2010 'blob' test).  A similar force expression was first proposed by
\citet{Ritchie2001}.  Geometric density averaged force expressions are now
employed in all modern SPH codes (e.g. \cite{Hopkins2013}, \cite{Saitoh2013},
\cite{Kawata2013}, \cite{Hu2014} and \cite{Read2010}).  As
stated in \cite{Read2010} and \cite{Saitoh2013}, a key requirement
for correct results with the modified force expression is a consistent energy
equation that conserves entropy (which we employ).  This is important to
correctly model strong shocks, such as Sedov blasts.  The extreme temperature
jumps at strong shocks also require the time-step limiter of \citet{Saitoh2009}.
The modern SPH code papers listed above all employ these updates and demonstrate
accurate solutions for strong shocks (e.g. Sedov blasts) and shear flows (e.g.
Kelvin Helmholz instabilities and the destruction of cold blobs).

For the tests shown here we used the Wendland $C_2$ kernel detailed in
\citet{Dehnen2012} with 64 neigbours where the SPH smoothing distance is defined
so that the kernel weight drops to zero at $2\,h$.  

Sharp density contrasts, as seen in highly resolved superbubbles, can require
additional checks on the neighbour finding component of the SPH method.  For
example, hot particles can inadvertently become hydrodynamically decoupled from
cold ones for a fixed number of neighbours.  A full set of cold neighbours can
sit at the edge of the kernel, where their contribution is negligible.  The hot
particle can thus have a full set of neighbours but feel minimal forces.  We
increase the number of neighbours until at least $18$ neighbours are within
$1.41\,h$.  

With respect to heating and cooling, for this work we used the
\citet{Ritchie2001} density when calculating cooling rates to sharpen the
density contrast between hot and cold gas.  This is particularly useful at lower
resolution when the hot bubble is resolved with a small number of particles.
This improves the ability of low resolution runs to give similar energy loss
rates to high resolution versions.
 
Feedback can increase gas particle masses substantially in the rare event that a
gas particle spends a lot of time within star clusters without forming a star
itself.  This can degrade the accuracy of the SPH method.  We avoid this problem
by splitting particles that exceed $4/3$ their initial mass into two equal mass
particles with the same properties.  This affects a very small fraction of the
particles.

These modifications, along with detailed testing, will be discussed in a
forthcoming paper on {\sc GASOLINE2}.  For reference, the quality of our results
on the tests discussed here is similar to the results presented for other modern
modified SPH codes \citep[e.g.][]{Hopkins2013}.   

\begin{figure}
    \includegraphics[natwidth=908, natheight=631, width=0.5\textwidth]{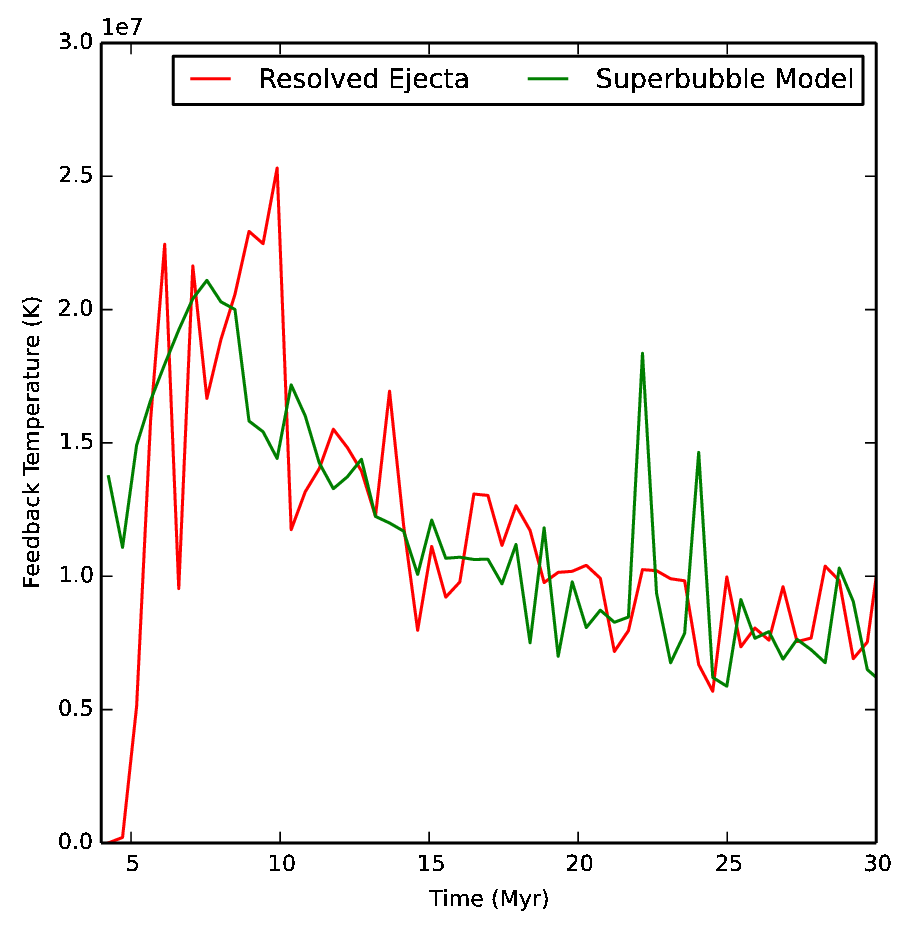}
    \caption{Peak temperature of feedback-heated gas inside the isolated star
        cluster's hot bubble.  Evaporation quickly enters the self-regulating
        regime, and the temperature is roughly constant for the last 
    $15\;\mathrm{Myr}$ of feedback.}
    \label{onestar_hottemp}
\end{figure}

\section{Simulations}
\subsection{High-Resolution Star Cluster Test}\label{cluster}
We began by exploring models of a single, isolated superbubbles at high
resolution.  For these tests, we employed three physical models.  The {\it
Direct injection} approach models as much physics as possible from first
principles.  Feedback mass and energy is modelled via a stream of new gas
particles created from the star cluster.  This approach only works when the gas
resolution elements are much smaller than the star cluster mass.  The only
component of this model that is sub-grid is evaporation as this occurs on
extremely small length scales.  {\it Superbubble Feedback} refers to our new
model.  The key addition over the direct approach is the sub-grid multiphase
treatment.  Feedback energy and mass is injected into existing particles which
may split.  

We also include a {\it Simple Feedback
Model} modeled after that proposed by \citet{Agertz2013}.  This model is a
stand-in for models typically used to date and does not include conduction or
evaporation.  Feedback mass and energy is given to the nearest gas particle to
the star cluster.  This simple model incorporates a two-component energy
treatment with radiative cooling disabled for feedback energy.  Feedback energy
is steadily converted into the regular, cooling form with an e-folding time of
$5\;\mathrm{Myr}$. 

\subsubsection{Basic Superbubble}

We placed a star cluster with mass $3\times10^4\;\mathrm{M_\odot}$ in an uniform
periodic box $2\times2\times2\;\mathrm{kpc}$ in size, containing
$10^3\;\mathrm{K}$ gas with solar metallicity at a density of
$1\;\mathrm{cm^{-3}}$.   This gives a gas particle mass of $760\; M_\odot$,
$6080\; M_\odot$, and $48640\; M_\odot$ for resolutions of $128^3$, $64^3$, and
$32^3$ respectively.  These tests do not include photoheating. 

Figure~\ref{onestar_column} shows the column density projection for the three
different feedback models in an isotropic medium.  injection model.   The hot
interior of the bubble produced using the simple feedback model contains less
mass at $128^3$ than the direct injection model and is subsequently too poorly
resolved for the instabilities formed in the accelerated shell
\citep{Vishniac1983} to mix the bubble and shell through turbulence and
diffusion.

As figure~\ref{onestar_homogeneous} shows, the superbubble feedback model
performs much better than the simple model in reproducing the hot mass
production rate from the high-resolution direct injection model, and lacks the
extreme resolution sensitivity that the simple model exhibits for the amount of
gas heated. The simple model only heats $\sim 4$ gas particles.  For the $128^3$
and $64^3$ resolutions, this gives too little hot mass.  For $32^3$ (extremely
poor resolution, in fact with gas particles more massive than the entire
cluster), this produces more than twice too much hot mass (more than
$2\times10^5\;M_\odot$). Meanwhile the simulation with the superbubble feedback
model only underestimates the hot mass by roughly a third of the target $128^3$
direct injection simulation when used at a resolution $32^3$.  The decreased
momentum in the lower-resolution runs is likely due to particles staying longer
in the multiphase state.  More massive particles take longer to fully evaporate,
and thus some mass that would be part of a pressure-driven cold shell at higher
resolution is tied up in the cold part of multiphase particles.

Figure~\ref{onestar_hottemp}  shows that the actual peak temperature of the
feedback-heated bubble in the isolated star cluster run is roughly
$1\times10^7\mathrm{K}$.

\begin{figure}
    \includegraphics[natwidth=947, natheight=358, width=0.5\textwidth]{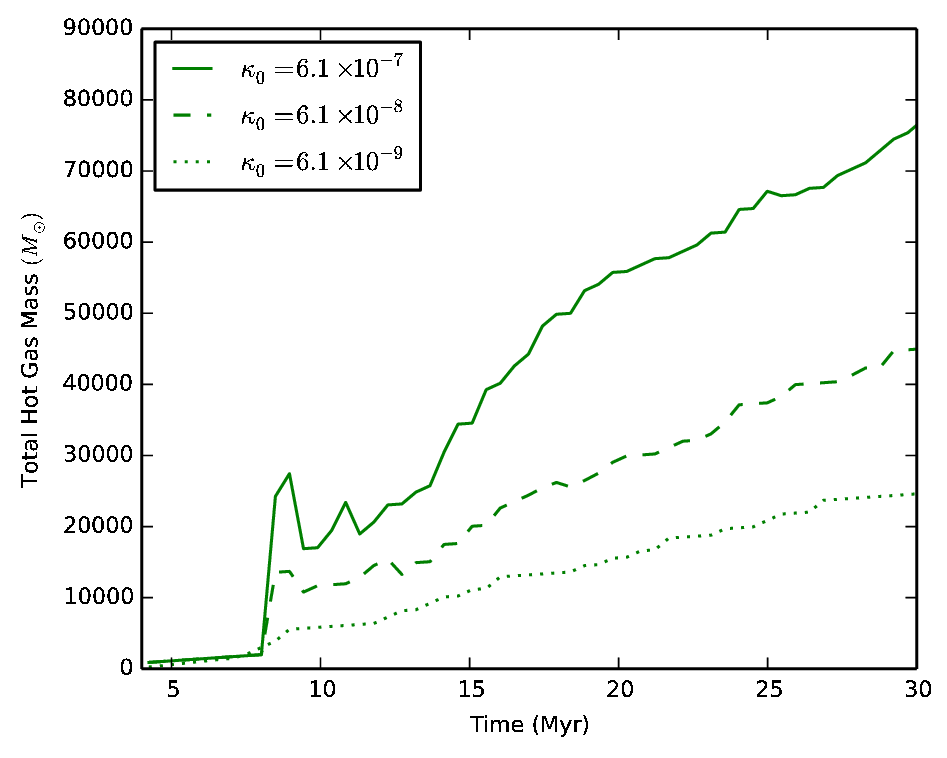}
    \caption{Hot mass production as a function of the conduction coefficient
        $\kappa_0$.  As this figure shows, reducing $\kappa_0$ by a factor of
        100 reduces the amount of hot mass generated through conduction by only
        a factor of $\sim2$. All $\kappa_0$ values have units of $\mathrm{erg\;
        s^{-1} K^{-7/2} cm^{-1}}$.}
    \label{onestar_kappa}
\end{figure}

\subsubsection{Suppressed conduction}

We also ran a set of simulations with the value of $\kappa_0$ used in the model
reduced by a factor of 10 and a factor of 100.  Since there is some uncertainty
as to this coefficient in a magnetized ISM, we use this test to show that the
self-regulating effect of conduction is insensitive to variation in $\kappa_0$.

Figure~\ref{onestar_kappa} shows the hot mass generated in simulations with 3
different values of $\kappa_0$.  The dashed curve shows a reasonable lower limit
for $\kappa_0$ \citep{Cowie1977}, while the dotted curve shows a much more
extreme conduction reduction than is expected in nature.  Both curves illustrate
the insensitivity of the method to reductions in the conduction rate.  Even
reducing $\kappa_0$ by 100 only results in a reduction of the hot mass inside
the bubble to just around a third.

\begin{figure}
    \includegraphics[natwidth=1972, natheight=969, width=0.5\textwidth]{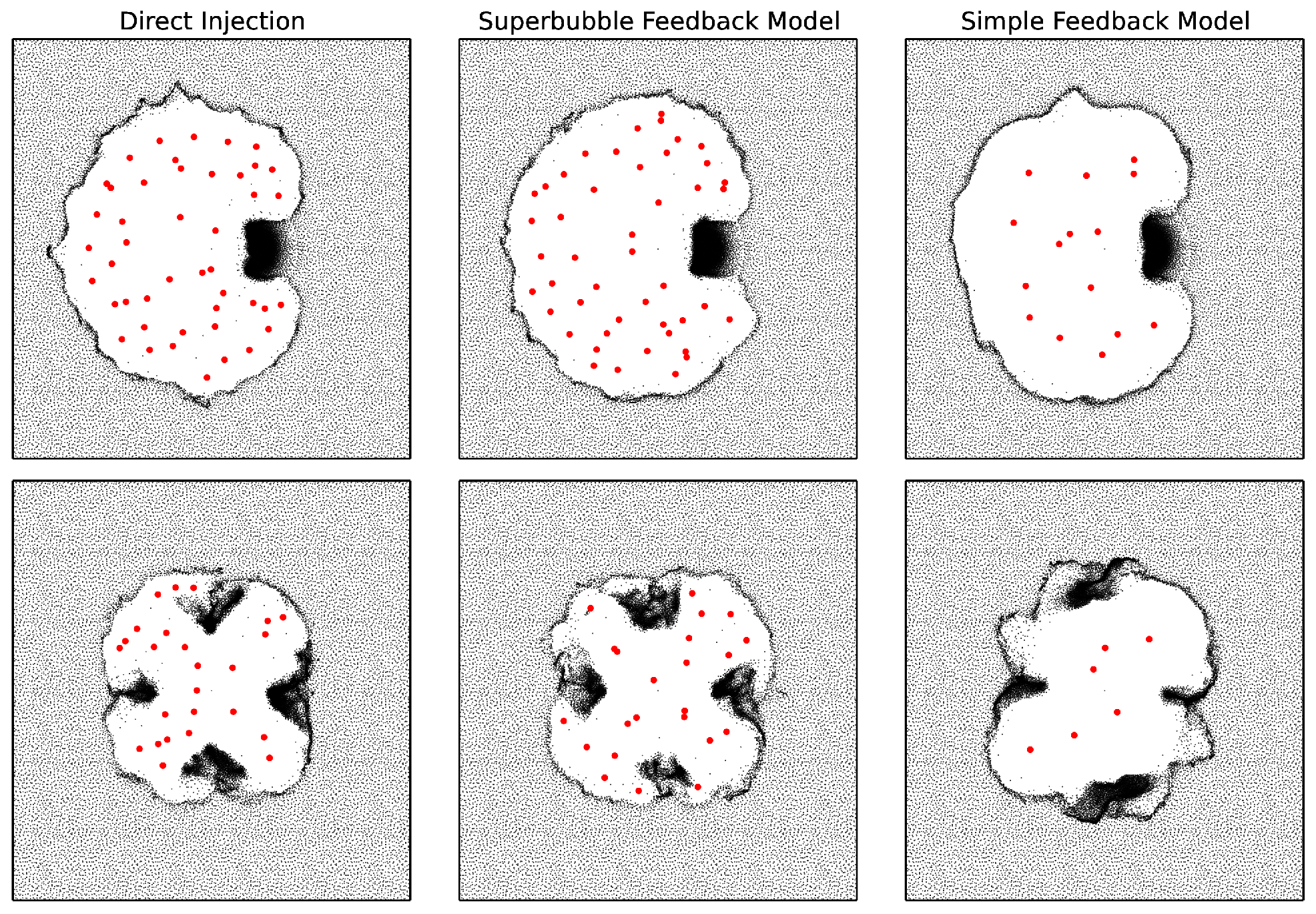}
    \caption{3 kpc wide slices from the same three methods shown in
        figure~\ref{onestar_column}, also at 50 Myr, but applied to a star
        cluster in an inhomogeneous medium.  Particles whose smoothing length
        intersects z=0~kpc are shown.  Particles above $10^5$ K are shown in red.
        The upper row shows simulations with a single clump, while the bottom row
        shows simulations with 6 clumps.}
    \label{onestar_clumpyslice}
\end{figure}

\begin{figure*}
    \includegraphics[natwidth=1833, natheight=631, width=\textwidth]{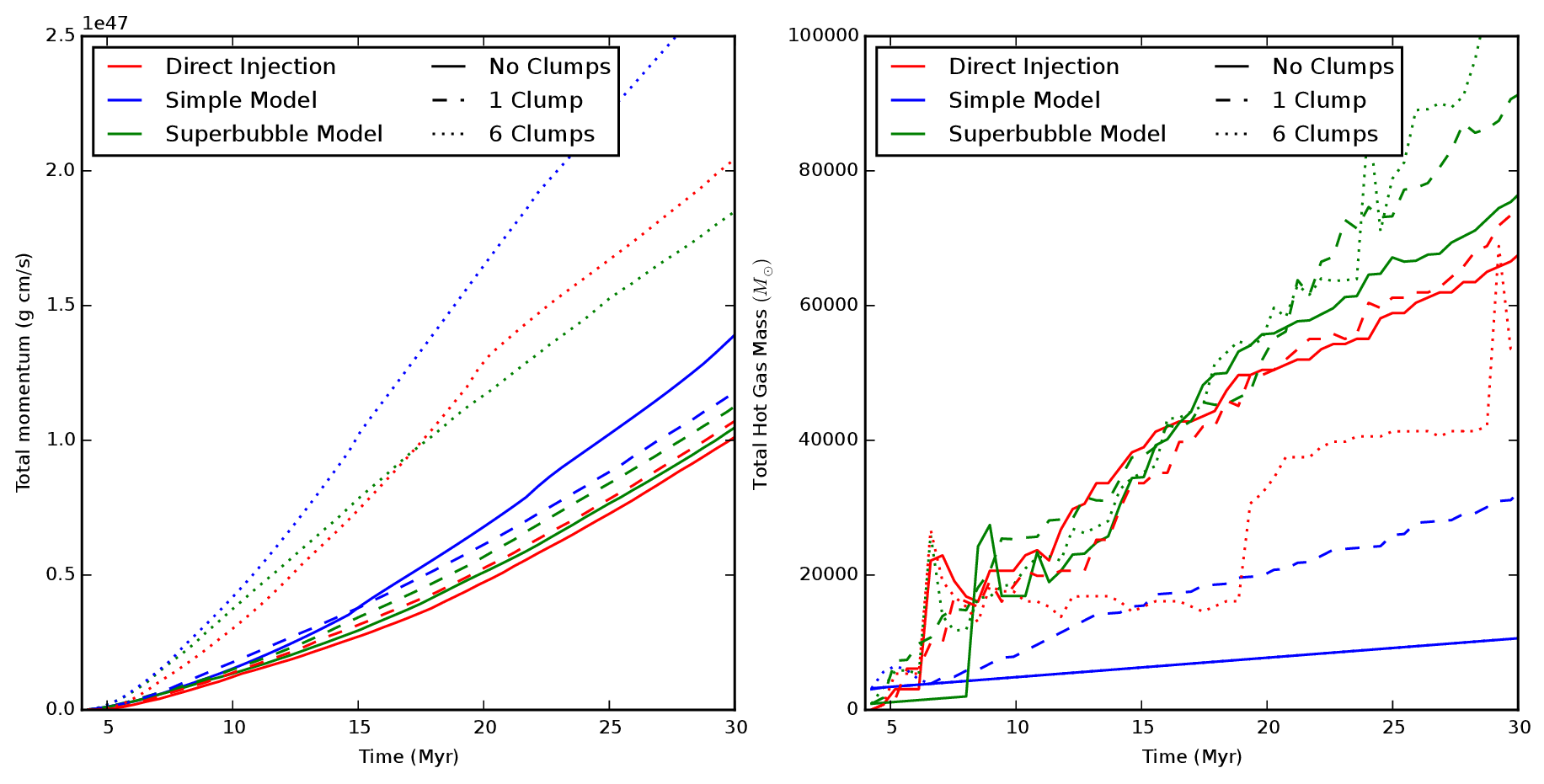}
    \caption{Feedback effects as a function of ISM homogeneity.  The above
        figure shows on the left the total radial momentum imparted to the
        medium, and on the right the total amount of mass heated to above
        $10^5\;\mathrm{K}$.  The red curves are the direct injection results, the
        blue corresponds to the simple feedback, while the green shows the
        results of the superbubble model.  Solid curves show the results for a
        homogeneous ISM, dashed for a single dense clump, and dotted for 6 dense
        clumps.}
    \label{onestar_clumpy}
\end{figure*}

\subsubsection{Clumpy Medium}\label{clumpy}
As the real ISM is highly inhomogeneous (and as \citet{Silich1996} showed, cold
clumps can also be a source of evaporated cold material) , we ran two additional
simulations with cold, dense clumps with $100\;\mathrm{cm^{-3}}$ density and
$10\;\mathrm{K}$ gas in an ambient medium of $0.5\;\mathrm{cm^{-3}}$ at
$1000\:\mathrm{K}$.  This gives roughly the same amount of mass enclosed within
the hot bubble at $30\:\mathrm{Myr}$.  The first contains a single spherical
clump with a radius of $0.2\;\mathrm{kpc}$.  The second contains the same amount
of cold mass spread over 6 clumps arranged at the center of each face of a cube
$0.2\;\mathrm{kpc}$ surrounding the star.  We use this idealized clumpy medium
to test the sensitivity of the model to small scale structure that may be
unresolved in lower resolution simulations. Figure~\ref{onestar_clumpyslice}
shows a slice showing SPH particles at $50\;\mathrm{Myr}$ for the two clumpy
cases.  

Figure~\ref{onestar_clumpy} shows that the superbubble model is also capable of
handling feedback into a clumpy medium.  With more than an order of magnitude
difference in the total hot mass compared to the simple model, along with less
sensitivity to environment or resolution, the superbubble model is better
suited for probing galactic outflows in simulations.

\begin{figure*}
    \includegraphics[natwidth=1856, natheight=319, width=\textwidth]{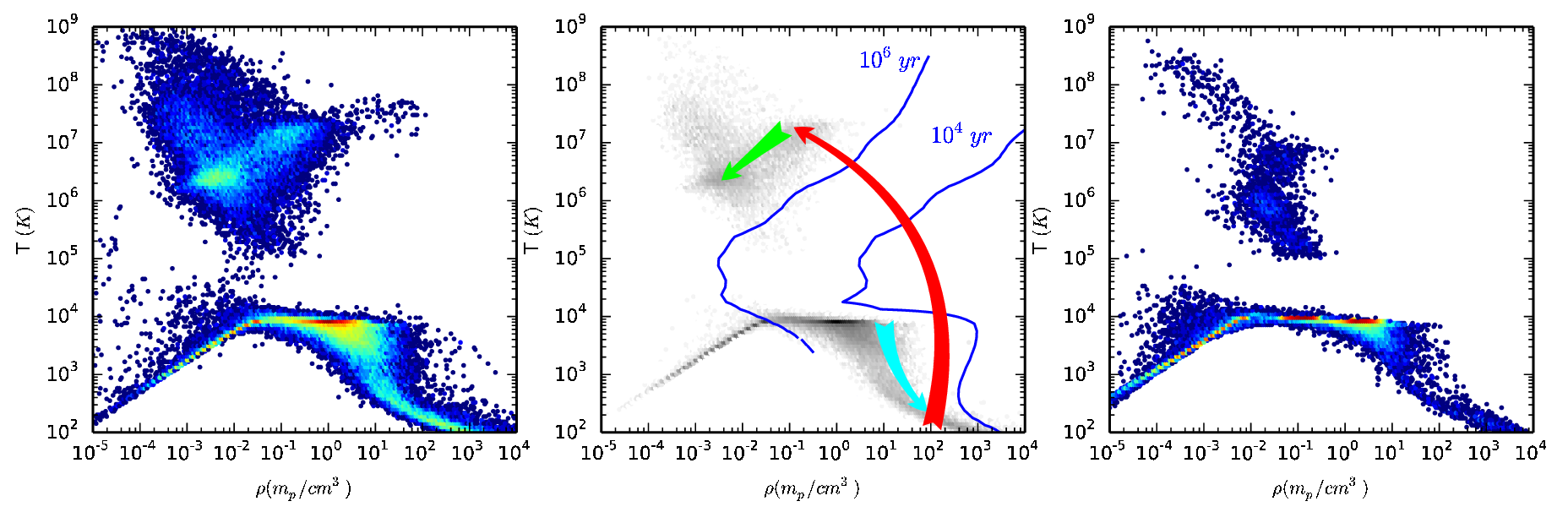}
    \caption{Phase diagrams for the Milky Way (left) and dwarf (right)
        simulations at $200\;\mathrm{Myr}$.  The central panel shows a typical
        path for gas ejected from the Milky Way.  First gas cools radiatively
        (cyan path) to high density, where it becomes multiphase as its hot
        component is heated to $\sim10^7\;\mathrm{K}$ by feedback (red path)
        from nearby stars formed from its neighbouring high density gas.  The
        hot phase cools primarily through adiabatic expansion (green path).
        This process is often repeated one or more times, with gas that is
        entirely hot phase being ejected from the disc.  Cooling times of
        $10^4\;\mathrm{yr}$ and $10^6\;\mathrm{yr}$ are shown in blue. The
        majority of the hot gas in the upper left quadrant of the phase
        diagram has cooling times $>10^8\;\mathrm{yr}$.  The cooling curve
        for $10^8\;\mathrm{yr}$ passes through the green curve.  Note that
        particles in the multiphase state show the temperature and density for
        each state as separate points.  The mean properties of multiphase particles
        would place them in regions with short cooling times.}
    \label{disc_phase}
\end{figure*}

\begin{figure*}
    \includegraphics[natwidth=1706, natheight=997, width=\textwidth]{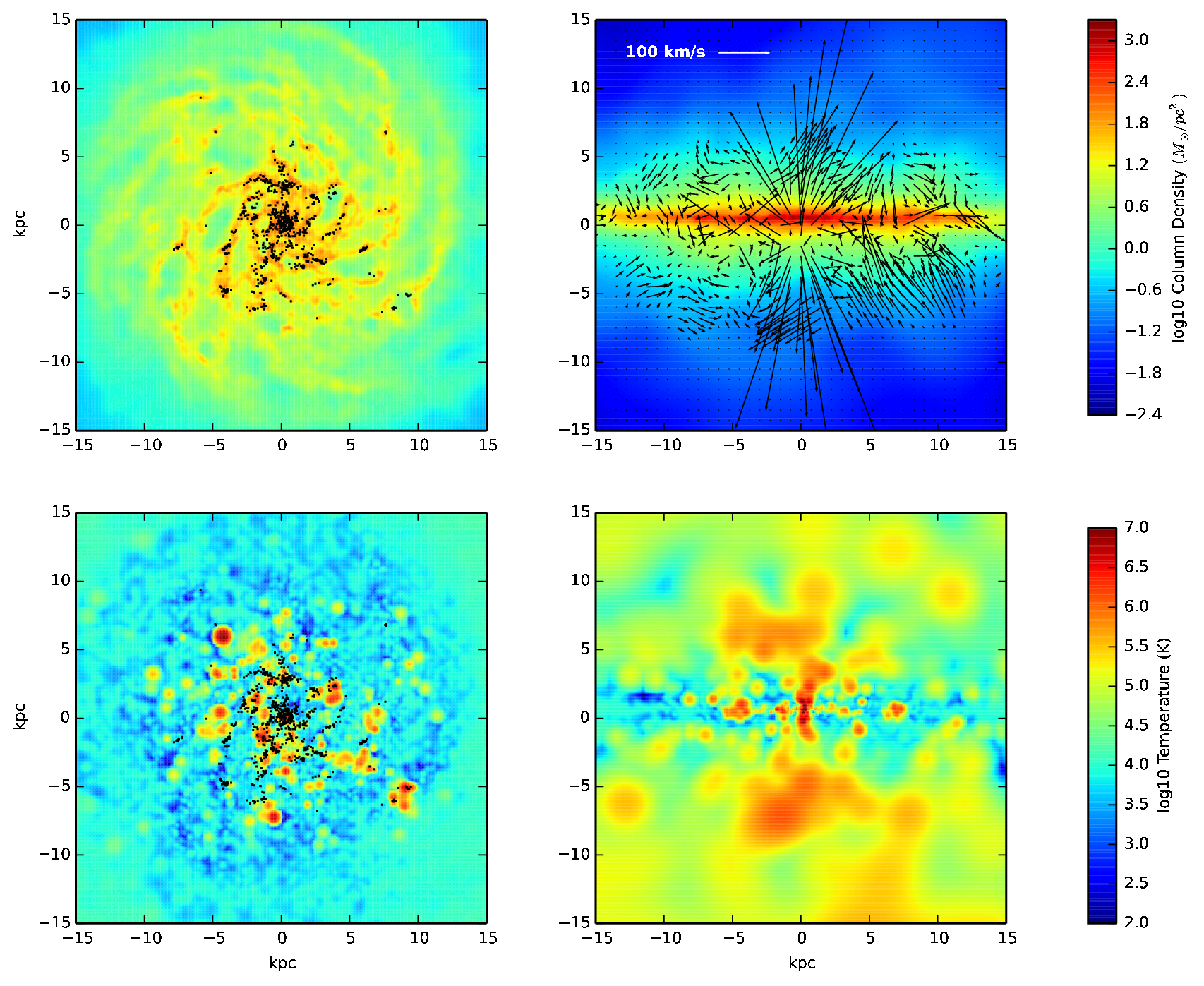}
    \caption{Column Density (upper row) and temperature for the Milky Way
        simulation at $300\;\mathrm{Myr}$.  The vectors show the in-plane velocity.
        Temperatures shown are averages between the two phases for multiphase
        particles.  Black points in the face-on images show stars formed within
        the last $20\;\mathrm{Myr}$. Note that gas is both leaving the disc near the galactic
        core, and returning in some places near the edge.}
    \label{milkyway_column}
\end{figure*}

\begin{figure*}
    \includegraphics[natwidth=1867, natheight=303, width=\textwidth]{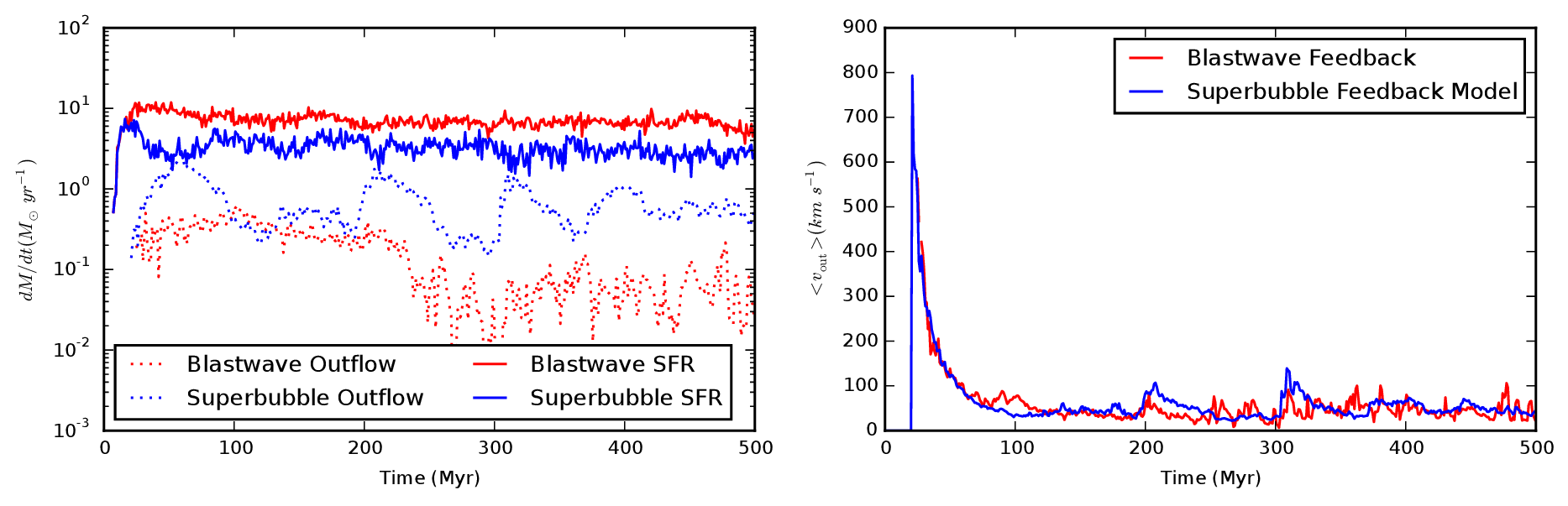}
    \caption{Outflow evolution for the Milky Way-like simulation.  The left
        hand plot shows the star formation rate and the outflow
        rate The right hand plot shows the average outflow
        velocity.}
    \label{milkyway_outflow}
\end{figure*}

\begin{figure*}
    \includegraphics[natwidth=1697, natheight=986, width=\textwidth]{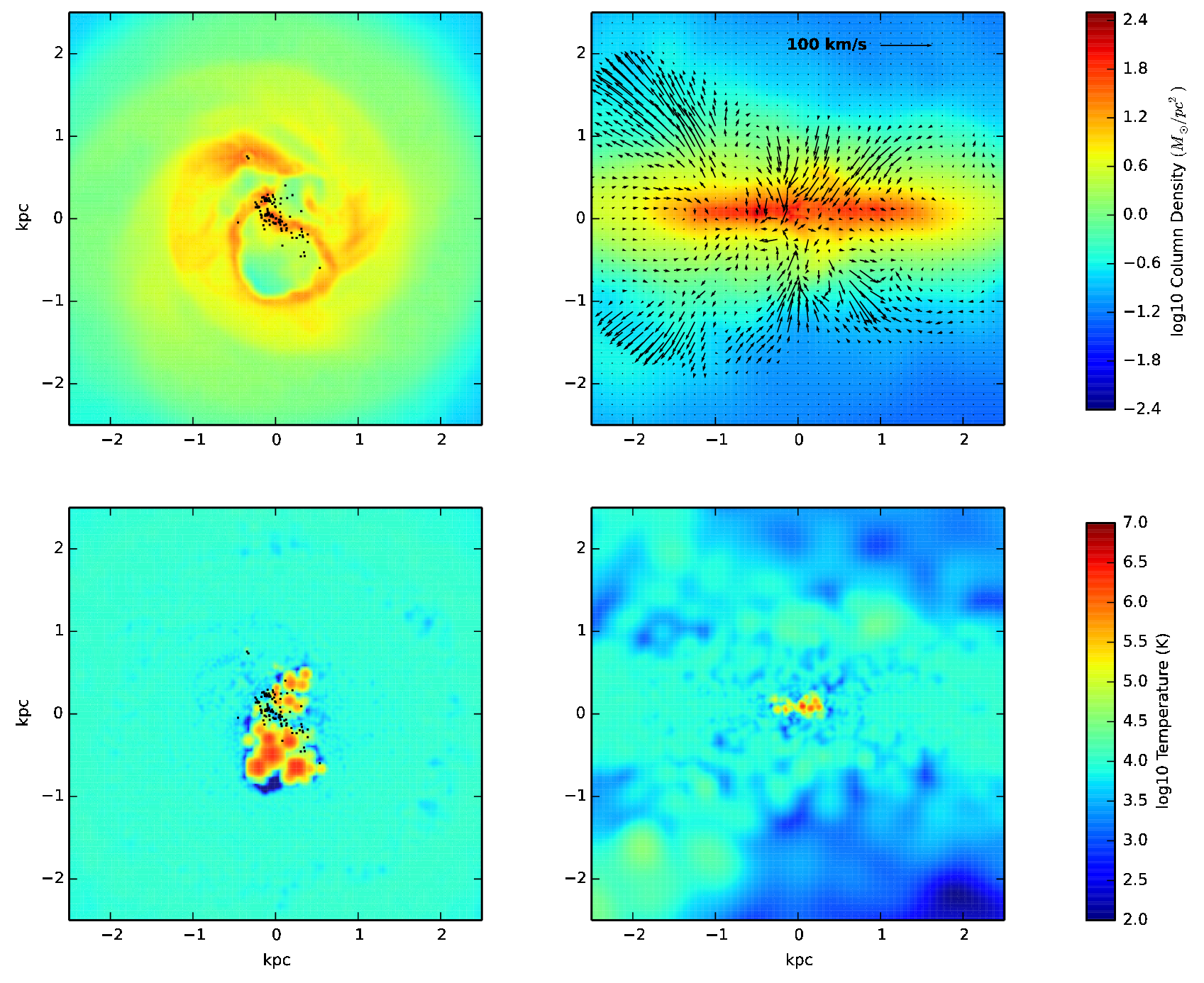}
    \caption{Column Density (upper row) and temperature for the dwarf simulation
        at $300\;\mathrm{Myr}$.  The vectors in show the in-plane velocity.  Temperatures
        shown are once again averages between the two phases for multiphase
        particles.  Note the much more `puffed up' appearance compared to the
        Milky Way, due to the more mass-loaded winds}
    \label{dwarf_column}
\end{figure*}

\begin{figure*}
    \includegraphics[natwidth=1867, natheight=403, width=\textwidth]{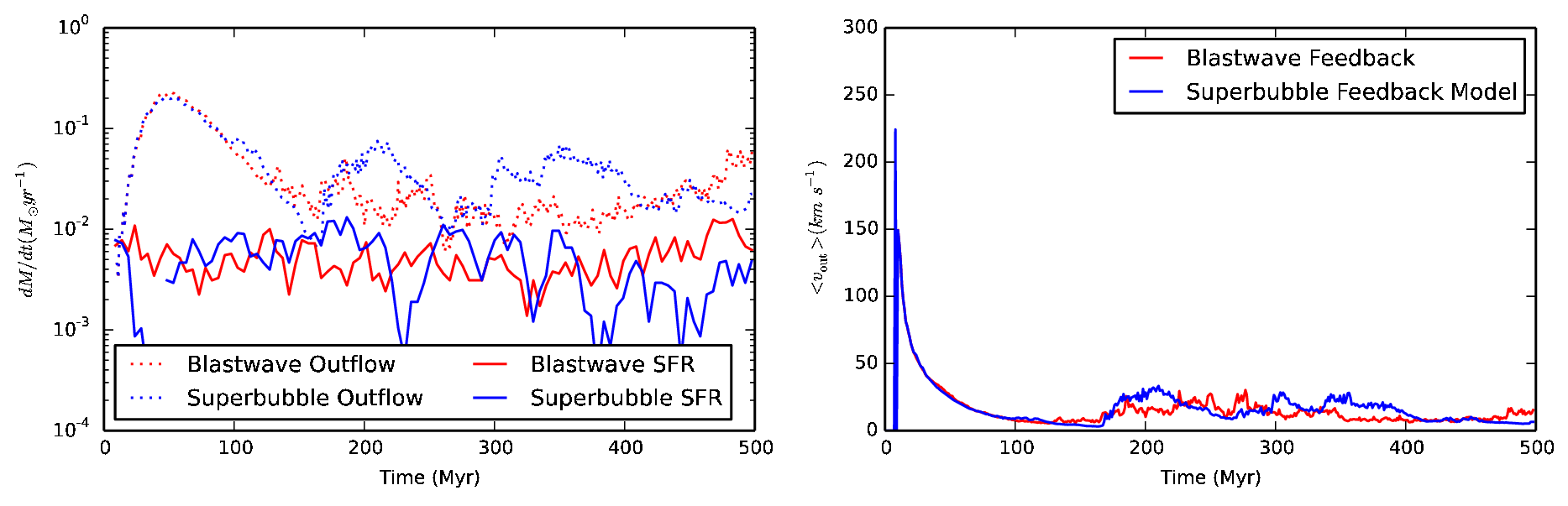}
    \caption{Outflow evolution for the dwarf galaxy.  Note that the vertical
    ranges are different than in figure~\ref{milkyway_outflow}}
    \label{dwarf_outflow}
\end{figure*}

\subsection{Galaxy Simulations}\label{galaxy}
\subsubsection{Initial Conditions and Parameters}
We used the isolated disc galaxy initial condition from the AGORA comparison
project \citep{Kim2013}.  These initial conditions were generated using the
equilibrium disc generating code of \citet{Springel2005}.
This galaxy is similar to a MW-type spiral galaxy
at $z=0$.  For our dwarf simulation, we scaled the masses down by a factor of
100, and the length scales by a factor of $100^{1/3}$, preserving the physical
densities in the initial conditions and lowering the surface density.  The dwarf
is thus similar to a low surface density local dwarf spiral.  These initial
conditions were intended to be similar to the $G10$ and $G12$ initial conditions
used in \citet{DallaVecchia2012}. The properties of these initial conditions are
shown in table 1.  Both simulations have 312500 total particles and 100000 gas
particles, so that the mass resolution is substantially higher in the dwarf.  The
initial gas metallicity is solar in both cases.

We used the standard {\sc GASOLINE} star-formation recipe, based on the
algorithm proposed by \citet{Katz1992} and detailed further in
\citet{Stinson2006}.  We use a density threshold for star formation $n_{SF}$
shown in table 1 along with a temperature threshold of $T < 10^4\;\mathrm{K}$.
Thus, for a given eligible gas particle, the probability of forming a star each
timestep $P_{SF} = 1-\exp{-0.05\ \Delta t\over t_{\rm ff}}$, depends only on the
free-fall time $t_{\rm ff}$.  This corresponds to the effective star formation
density rate of $\dot \rho_* = 0.05\ \frac{\rho_{gas}}{t_{\rm ff}}$.  We also
include UV heating for $z=0$ \citep[as in][]{Shen2010} and a pressure floor that
ensures gas does not collapse beyond the resolvable Jeans length
\citep{Machacek2001}.  

We also simulated these initial conditions using the established `blastwave'
feedback model from \citep{Stinson2006}, which has been a standard feedback
model for galaxy simulations in numerous previous studies.
\begin{table}
    \begin{tabular}{ c c c c c}
        \hline
        Simulation & $M_{tot} \mathrm{(M_\odot)}$ & $M_{gas} \mathrm{(M_\odot)}$
        & $\epsilon \mathrm{(pc)}$ & $n_{SF} \mathrm{cm^{-3}}$\\
        \hline
        Milky Way & $1.3\times10^{12}$ & $8.6\times10^9$ & 20  & $>10$\\
        Dwarf & $1.3\times10^{10}$ &  $8.6\times10^7$ & 4.3 & $>50$\\
    \end{tabular}
    \caption{Disc galaxy initial conditions.  $\epsilon$ is the gravitational softening and $n_{SF}$ is the star formation threshold.}
\end{table}

\subsubsection{ISM properties and Star Formation Rates}\label{ISM}

\begin{figure}
    \includegraphics[natwidth=942, natheight=642, width=0.5\textwidth]{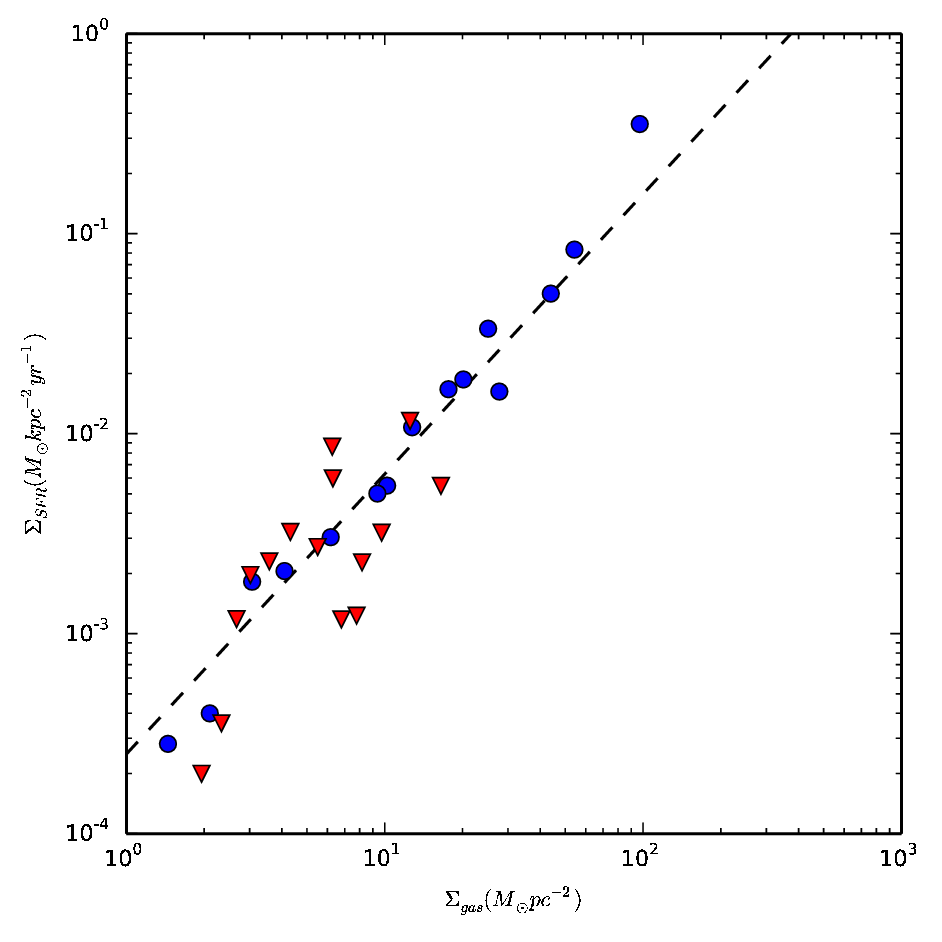}
    \caption{Kennicutt-Schmidt law in the Milky Way-like (blue points) and dwarf
        (red triangles) galaxies at $500\;\mathrm{Myr}$. Surface densities were
        calculated in radial annuli.  The dashed line shows the
        Kennicutt-Schmidt Law.  The superbubble model is easily able to regulate
        star formation rates to within this range.}
    \label{kennicutt_schmidt}
\end{figure}

\begin{figure}
    \includegraphics[natwidth=958, natheight=619, width=0.5\textwidth]{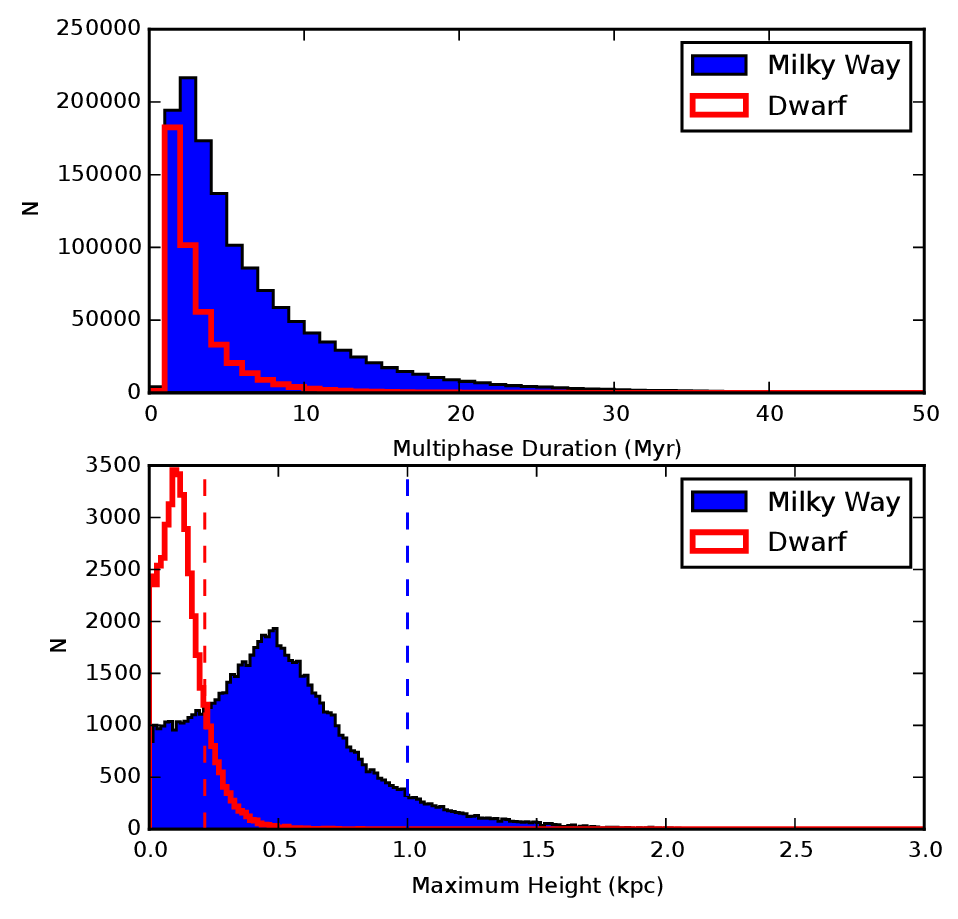}
    \caption{Properties of multiphase particles in the galaxy simulations.  The
        top figure shows duration of multiphase state for particles in both
        galaxy simulations.  Particles that are in the multiphase state more
        than once (convert back to single phase, cool, and then receive feedback
        again) have each time they are multiphase counted separately. The bottom
        figure shows maximum heights reached by particles in the multiphase
        state.  For each particle that is ever in the multiphase state, the
    maximum height it reaches {\it while still multiphase} is shown above for
    both galaxy simulations.  Dashed lines show scale heights at
    $10^4\;\mathrm{K}$ for each
of the two simulations.}
    \label{multiphase_properties}
\end{figure}

Figure~\ref{disc_phase} shows phase diagrams from both Milky Way and dwarf
galaxies.  Figure~\ref{milkyway_column} shows column density and temperature for
the Milky Way, and figure~\ref{milkyway_outflow} shows star formation rates and
outflow properties.  Figures~\ref{dwarf_column}~and~\ref{dwarf_outflow} show the
same for the dwarf galaxy.  The Kennicutt-Schmidt relation for both galaxies is
shown in figure~\ref{kennicutt_schmidt}.  The star formation and outflow properties are
discussed in section~\ref{Morphology}.  Finally, we show some properties of the
multiphase particles in figure~\ref{multiphase_properties}.

Figure~\ref{disc_phase} shows that phase diagrams for galaxies using superbubble
feedback strongly distinguish between pre- and post-feedback gas.  Above $\sim
10^5\;\mathrm{K}$, we see (especially in the Milky Way), a hot medium of
including halo gas and low density gas inside superbubbles within the galaxy
disc (see the temperature slices in
figures~\ref{milkyway_column}~and~\ref{dwarf_column} for images of the gas
temperature in these bubbles).  The bulk of the gas lies at a roughly $10^4\;
\mathrm{K}$ equilibrium between cooling and photoheating from the UV background.
A cold medium of both dense shells surrounding superbubbles and cooling clouds
(soon to form stars) also forms in both the Milky Way and (to a lesser extent)
the dwarf.  The central panel shows a schematic of how gas migrates between
these regions.  Radiative cooling (blue arrow), bring gas to high densities.
Feedback creates a second hot phase in nearby gas particles.  The cold component
is relatively unaffected though it can compress due to the increased pressure
(staying near the tip of the blue arrow).  The hot and cold phases of multiphase
particles are plotted separately.  The hot component immediately moves to low
density and high temperatures,  $\gta 10^7$ K (the red arrow).  If the particle
continues to receive feedback, evaporation rapidly consumes its cold part and
the particle can flow out to the halo and remain buoyant and slow cooling.  It
evolves adiabatically as it does so (green arrow).

This panel is telling in that it shows that no gas is found within the
`forbidden' region of short cooling times of $\lta 10^4\;\mathrm{yr}$.
Cooling-shutoff methods often produce large quantities of this gas in a high
temperature, high density state.  If we were to simply take the average
temperature and density of the multiphase particles, they would almost entirely
lie within this region, on the line connecting the cold and hot phases (which,
of course, is exactly the impetus for using multiphase particles, since a
particle with the average properties would cool away all of its feedback energy
much too rapidly).

The roughly fixed amount of gas heated in the superbubble simulations shown
previously gives a roughly constant feedback-heated gas temperature of $\sim
2\times10^7\;\mathrm{K}$.  Figure~\ref{onestar_hottemp}  shows that the actual
peak temperature of the feedback-heated bubble in the isolated star cluster run
varies between slightly less than this, to $\sim 1\times10^7\mathrm{K}$,
due to some cooling in the hot bubble.  This suggests that the model should
behave similarly to the stochastic thermal feedback model presented in
\citet{DallaVecchia2012}.  

As the multiphase fluid particles exist to bridge the gap between when the hot
interior of a superbubble contains too little mass to be resolved and the later
stage when resolved physics can take over, we should find that particles stay in
this phase for only a fraction of the lifetime of a superbubble.  From
\citet{MacLow1988}, the cooling time for superbubbles is on the order of a few
$10\;\mathrm{Myr}$, with a weak dependence on feedback luminosity and the
surrounding ISM density and metallicity.

We should expect that on average, multiphase particles convert back to single
phase in less than this time, a few $\mathrm{Myr}$.  In addition to their
lifetimes, we should expect multiphase particles to cluster in the discs of our
galaxy simulations (since they are spatially correlated with the stars that are
heating them), and that hot winds leaving the galaxy are composed of
fully-resolved, hot gas.  These winds are released when superbubbles grow large
enough to break out of the denser disc ISM, and thus should be well within the
resolved phase for these simulations.

Figure~\ref{multiphase_properties} shows the that all particles stay in
the multiphase state as well as the maximum height reached by multiphase
particles.  The top figure shows that the vast majority of particles in either
the dwarf or the Milky Way convert back to single phase within
$10\;\mathrm{Myr}$.  The mean multiphase lifetime for the Milky Way was
$6.6\;\mathrm{Myr}$, and $2.7\;\mathrm{Myr}$ for the dwarf, well within the
range we should expect.  The reason for the shorter multiphase lifetimes in the
dwarf galaxy is simply the better mass resolution: the hot bubble interior
becomes resolved earlier in the dwarf than in the Milky Way.  Fewer than 0.5 per
cent of multiphase particles ever stay in the multiphase state for more than
$50\;\mathrm{Myr}$ in both the Milky Way and the dwarf.  The bottom figure
shows, again as we should expect, that most multiphase particles convert back to
single phase before they reach 1 scale height.  The mean maximum height for
multiphase particles was roughly $1/2$ a scale height for both simulations,
$0.51\;\mathrm{kpc}$ and $0.13\;\mathrm{kpc}$ for the Milky Way and dwarf
respectively.  Neither simulation had any multiphase particles reaching heights
of more than $10\;\mathrm{kpc}$ before converting back to single phase.  This
shows that multiphase particles are essentially embedded within the thin disc of
the ISM: all of the mass outflowing is fully-resolved hot gas, and its behaviour
is fully governed by standard hydrodynamics.  The winds driven from both
galaxies are ejected through nothing more than simple buoyancy.

\subsubsection{Galaxy Morphology and Outflows}\label{Morphology}

In order to compare superbubble feedback to \citet{DallaVecchia2012}, we selected
similar galaxies to the ones shown in their paper and calculated the properties
of their outflows using a similar method.  We adopt their two primary metrics,
mass outflow rate $\dot M$ and mean outflow velocity $<v_{\rm out}>$.  These two
metrics give us an idea as to both how much gas is ejected from the galaxies,
and how long that gas will take to return to the disc from the halo (if it does
return at all).

We calculated outflows from our galaxy simulations by selecting particles that
are moving away from the galaxy between a planar region 5 scale heights (5kpc
for the Milky Way, 1.13kpc for the dwarf) above and below the disc, and 0.5
scale heights thick.  The outflow rate $\dot M$ is simply the total momentum of
outflowing particles (particles returning on fountains are excluded) passing
through this region divided by the thickness of the region.  The average outflow
velocity $<v_{\rm out}>$ is just the mean velocity of these same particles.

The superbubble feedback method, as shown in figure~\ref{kennicutt_schmidt} (and
the SFR shown in figures~\ref{milkyway_outflow}~and~\ref{dwarf_outflow}), also
passes the most basic requirement for a useful feedback model:  it is capable of
regulating star formation to match observed global star formation efficiencies.
This result is a typical outcome for effective feedback models with density-based 
star formation rates \citep[e.g.][]{Springel2003}.
Note that the simple star formation prescription used for
these tests does not have a cut-off at lower surface densities.  It is clear in
figure~\ref{milkyway_outflow} that the average SFR in the Milky Way is roughly
twice as high in the simulation using blastwave feedback as compared to the
simulation with superbubble feedback.  It is also clear from this figure that
outflows in the Milky Way simulation with superbubble feedback contain
approximately ten times the mass of outflows driven by blastwave feedback.

\section{Discussion}

It is important to heat the right amount of gas through feedback. This is
particularly important if one wishes to examine feedback-driven galactic winds.
As figures~\ref{onestar_homogeneous}~and~\ref{onestar_clumpy} show, the simple
cooling-shutoff feedback model produces quite different amounts of hot gas as a
function of both resolution and ISM homogeneity.  If one underestimates the
amount of mass heated by feedback, the winds one drives will be hotter, but
contain less mass.  In other words, outflows will be faster but contain less
mass.  If one overestimates the amount of mass, outflows may carry a larger
fraction of the galaxy's gas mass, but will be less able to actually remove
this gas from the galaxy, either permanently or for a long cycling timescale.
Either error will have  serious implications for predictions of the effects of
outflows on both host galaxies and the intergalactic medium.  We have
constructed our model such that both the momentum and the amount of hot gas
within a superbubble are resolution independent.  This is confirmed in
figure~\ref{onestar_homogeneous} over a range of mass 
resolutions from $\sim95M_\odot$ to $5\times10^4M_\odot$.  
Even in the extreme limit of a one particle superbubble,
the results are qualitatively correct and vary less than a factor of 2
from the expected solution.

In fact, for any feedback model that omits thermal conduction or other mixing
between the hot interior of a feedback bubble and the surrounding cold shell,
the amount of hot mass produced will be set by the resolution of the simulation
alone.  In fact, figure~\ref{onestar_homogeneous} shows this quite clearly.  For
each resolution, the hot mass produced is roughly constant, simply a product of
the simulation mass resolution and the number of particles feedback is shared
with.  Changing either of these will drastically change the amount of hot gas
generated by feedback.  

In general, the gas driven in outflows from these galaxies does not move fast
enough to escape the galactic halos in either the dwarf and the Milky Way.  This
is reasonable for star formation rates well below the starburst regime.  The
majority of gas ejected from both simulations instead cycles between the halo
and the disc.  This helps moderate star formation in the disc.  By
$300\;\mathrm{Myr}$, only $\sim 1.0$ per cent, or roughly
$8.4\times10^7\;\mathrm{M_\odot}$, of the Milky Way gas has been lifted to above
5 scale heights while the dwarf cycles more than a third of its total gas mass,
$3.3\times10^7\; \mathrm{M_\odot}$ into a fountain above 5 scale heights.  In
both simulations, this cycling induces periodic bursts of star formation and
disc outflows.

As figures~\ref{milkyway_outflow}~and~\ref{dwarf_outflow} show, the superbubble
method results in galaxies with stronger star formation regulation, and
more mass-loaded outflows than the well-established blastwave model.  
For example, the simulation of the Milky Way analog, the star
formation rate is lower by a roughly a factor of 2 compared to the blastwave (a
point in favour of the superbubble model, since \citet{Scannapieco2012} showed
that a cosmological simulation of the Milky Way using blastwave feedback in
{\sc GASOLINE} formed stars at roughly twice the rates observed by
\citet{Guo2011}).  We also see roughly an order of magnitude more gas ejected
from the disc with the superbubble model, as we expect from the results of the
single star cluster test, since the production of more hot mass should result in
more mass-loaded winds.  Interestingly, the mean outflow velocity shows only
small differences between the superbubble and blastwave models.  This means that
even though the outflows driven by the blastwave contain less mass, they do not
leave the galaxy any faster, and are no more likely to escape the galactic
potential than winds produced in the superbubble simulations.

Our superbubble model is comparable to the model of \citet{DallaVecchia2012}
with a $\Delta T \sim 10^{7}\;\mathrm{K}$, somewhat smaller than their fiducial
value.  Both heat of order $300\;\mathrm{M_\odot}$ per SNe.
\citet{DallaVecchia2012} simply relied on a stochastic model to deposit enough
energy only when a specified temperature can be reached.   Their model suffers
from overcooling at cosmological resolutions with densities $n_H >
10\;\mathrm{cm^{-3}}$ as their stochastic model requires the heating of
fractional gas particles to yield the temperatures they desire.   At moderate
resolution, some star forming regions thus experience no feedback and others get
strong feedback.  The multiphase mechanism can handle resolutions where the
initial feedback-heated gas mass is less than a single resolution element,
without relying on stochastic feedback.

The superbubble method has a number of distinct advantages over previous feedback
methods (cooling shutoffs, constant-temperature stochastic feedback,
hydrodynamic decoupling, etc.).  The superbubble model introduces no additional
free parameters, requiring only the total stellar feedback energy to be
specified.  The superbubble paradigm can incorporate multiple sources of
mechanical luminosity, primarily stellar winds and supernovae, within a single
framework.  Unlike other methods, the amount of mass heated by feedback is
physically motivated: it is the amount of mass evaporated into the hot bubble
through thermal conduction.  Radiative cooling is suppressed not by simply
disabling it (which at best only approximates the long cooling times desired),
but by injecting energy into a distinct low density, hot phase.  This allows the
model to handle high-resolution isolated galaxy simulations as well as
lower-resolution cosmological ones.  

Another distinct advantage of this method is that it is {\it local}.
Feedback-heated gas needs no knowledge of its environment save the information
it has already through hydrodynamic interactions with its neighbours.  This
allows the method to handle feedback from clustered star formation without any
additional changes; gas particles need not know the {\it total} energy inside
a feedback-heated superbubble, which can be difficult to determine in a bubble
that is heated by multiple stars and contains many gas particles or cells.
Clustered star formation is an important aspect of galactic evolution, and can
amplify the effects of feedback by concentrating it on a single region
(something done by hand in stochastic, constant temperature feedback models).
As resolution in galaxy simulations has been steadily improving, well-resolved
clustered star formation is beginning to become a reasonable goal and object of
study.  The superbubble model will allow feedback from these clusters to behave
in a physically correct manner that is insensitive to resolution.

\subsection{Summary}
Stars preferentially form in clusters.  Clustered stellar feedback generates
superbubbles which are qualitatively different to isolated supernovae.
Correctly evolving superbubbles requires the inclusion of thermal conduction.
Thermal conduction, acting on very small length scales, evaporates cold material
into hot feedback bubbles which must be captured via a sub-grid evaporation
model such as the one presented here.  At the typical resolutions achievable in
galaxy simulations, a sub-grid multiphase treatment is required to accurately
follow the evolution of the hot phase.  Combining these elements results in a
feedback model with several attractive features:

\begin{enumerate}
    \item Separate hot and cold phases within an unresolved superbubble prevent overcooling
        without relying on ad-hoc cooling shutoffs
    \item Feedback from multiple sources (e.g. star clusters) is combined correctly
    \item Feedback gas doesn't unphysically persist in phases with extremely short cooling times
    \item Star formation is strongly regulated: at least as effectively as current models with the same
           feedback energy
    \item The feedback can effectively drive outflows
    \item Evaporation due to thermal conduction
        generates the correct amount of hot gas which 
        subsequently determines galactic wind mass-loading
    \item For well resolved bubbles, the model no longer relies on its multiphase component 
         and naturally produces superbubbles that behave as predicted
    \item The model is insensitive to resolution
\end{enumerate}

\section*{Acknowledgements}
The analysis was performed using the pynbody package
(\texttt{https://github.com/pynbody/pynbody}, \citep{pynbody})
  We thank Mordecai-Mark Mac Low for useful
conversations regarding this paper.  The simulations were performed on the
clusters hosted on {\sc sharcnet}, part of ComputeCanada.  We greatly
appreciate the contributions of these computing allocations. James Wadsley and
Hugh Couchman thank NSERC for funding support.
\bibliographystyle{mn2e}
\bibliography{references}

\clearpage

\end{document}